\documentclass[structabstract]{aa}
\usepackage[dvips]{graphicx}
\usepackage{txfonts}

\def\be{\begin{equation}}
\def\ee{\end{equation}}
\def\bfi{\begin{figure}}
\def\efi{\end{figure}}
\def\bea{\begin{eqnarray}}
\def\eea{\end{eqnarray}}
\begin{document}
\voffset=+1truecm
\title{Time-energy correlations in  solar flare occurrence}
\author{E. Lippiello$^{a}$ \and L. de Arcangelis $^{b,c}$ \and C. Godano$^{a}$}
\institute{ $^{a}$ Department of Environmental Sciences and CNISM,
Second University of Naples, 81100 Caserta, Italy\\
$^{b}$ Department of Information Engineering and CNISM, 
Second University of Naples, 81031 Aversa (CE), Italy\\ 
$^{c}$ Institute Computational Physics for Engineering Materials, ETH, Z\"urich (CH)
}
\abstract
{The existence of time-energy correlations in flare 
occurrence is still an open and much debated problem.}
{This study addresses the question whether 
statistically significant correlations are present between  energies
of successive flares as well as energies  and  waiting times.}
{ We analyze the GOES catalog with a statistical approach
based on the comparison of the real catalog with a reshuffled
one where energies are decorrelated. This analysis 
reduces the effect of background activity and is able to reveal the role of
obscuration.} { We show the existence of non-trivial correlations between 
waiting times and energies, as well as between energies of subsequent flares.
More precisely, we find that flares close in time tend to have the second event
with large energy. Moreover, after large flares  the flaring rate
significantly increases, together with  the probability of other large flares.}
{Results suggest that correlations between energies and waiting times 
are a physical property and not an effect of obscuration.
These findings could
give important information on the mechanisms for energy storage and
release in the solar corona. } 

% These findings could
%give important information on the mechanisms for energy storage and
%release in the solar corona. } 

%\end{abstract}
%\pacs{96.60.Rd, 89.75.Da, 64.60.Ht}

\maketitle

\section{Introduction}

Solar flares are violent explosions of magnetic energy in the solar corona.
The theory of magnetic reconnection (see Priest \& Forbes \cite{Pri}
for a review) represents  the most plausible
and widely accepted explanation for flare occurrence, although
a definite and clear explanation of the mechanisms at the basis of
flare triggering is still lacking. 
A well-established property of flare occurrence is the power law decay
of the  peak-flux energy distribution (Lee et al \cite{Lee};
Aschwanden et al \cite{Asc}; Crosby et al  \cite{Cro}),
a property shared by
other stochastic physical phenomena like earthquakes (de Arcangelis et al \cite{deA}).
%with an exponent that is quite independent on the solar cycle phase and on the
%energy range considered. (Ashwanden, Crosby..)
Several mechanisms have been proposed to reproduce the above
experimental findings.
In the Rosner and Vaiana (RV) model (Rosner \& Vaiana \cite{Ros}) it is assumed that 
flare occurrence is an uncorrelated Poisson process where the energy
grows  at a rate  proportional to the internal energy of the system.   
Given this assumption, the energy storage 
follows an exponential temporal growth, interrupted at random when
the system releases all 
the stored energy in a flare. This mechanism accounts for the power law in the 
size distribution 
and predicts correlations between the storage time and the released energy. 
More precisely,
the later a flare will occur the higher the energy will be. 
Also the avalanche model (Lu \& Hamilton \cite{Lu}; Hamon et al  \cite{Ham}), 
which describes solar flares as energy 
relaxation events in a system driven at a constant rate, correctly predicts 
scale-free behavior for the peak-flux energy 
distribution. Within this  approach, flare occurrence is again a
Poisson process,
as in the RV model,  but occurrence times and energies are
uncorrelated like in standard self-organized models (Jensen \cite{jen}).
 Finally, power law behavior for the size distribution is also
consistent with models assuming correlations between the waiting
times $\Delta t$ of successive bursts, as in a shell model of 
magneto-hydrodynamic turbulence (Boffetta et al \cite{Bof}). 
This model reproduces the  experimental power law decay    
of the waiting time distribution  $P(\Delta t)$ which can 
also be obtained however by means of an uncorrelated piecewise Poisson process 
(Wheatland et al \cite{Whe3}; Norman et al \cite{Nor}). 

The investigation  of  
correlations between flare energies and waiting times is a useful tool to 
distinguish among different triggering mechanisms. 
This problem has been addressed in a series of papers
(Crosby et al  \cite{Cro}; Wheatland et al \cite{Whe1}; Wheatland \cite{Whe2}), 
providing no clear evidence for 
time-energy correlations. 
In particular, Wheatland et al. (\cite{Whe1})
detected small correlations at short time and interpreted them as
a spurious effect of obscuration, i.e. long-lasting events may hide
subsequent  small  events. By properly taking into
account obscuration effects, Wheatland concluded that no significant
correlations are present.
In this paper we present a novel analysis of experimental catalogs
providing  evidence for time-energy correlations
in flare occurrence. 
%These  findings suggest possible 
%mechanisms for energy storage and release in the flaring process. 
%We implement 
%these mechanisms in a stochastic model for flare occurrence and reproduce the
%experimental observations.  

\section{Data analysis and methods}

In our analysis  we used data from the GOES (Geostationary
Operational Environmental Satellite)  catalog 
(ftp://ftp.ngdc.noaa.gov/STP/SOLAR\_DATA/SOLAR\_FLARES) for the
solar cycles 21, 22 and 23 in the years 1977 to 2007.
We include only C1.4
class flares in the analysis (peak flux energy larger than  $E_0=1.4 \times 10^{-6}$
$\rm{W m^{-2}}$). Each flare in the catalog is characterized by its starting
time $t_i$ and its peak-flux energy $E_i$, 
%We define the flare magnitude $m_i =\log(E_i/E_0)$, 
in the following called "energy" to simplify notation and  measured
in unit of $\rm{W m^{-2}}$, counting $38575$ flares with $E \ge E_0$.
In order to enhance the eventual presence of temporal and energy
correlations 
among flares, we performed a recently proposed (Lippiello et al \cite{Lip1})
statistical analysis of the catalog.
We computed the probability to have a given quantity $x_i$ greater than $X$
conditioned to $\Delta t_i$ being smaller than $T$,   $P(x_i>X \vert \Delta t_i
<T)$. Here, $\Delta t_i=t_i-t_{i-1}$ is the waiting time 
and the quantity $x_i$ represents, depending on cases, the energy
of the $i$-th flare $x_i=E_i$, the energy of the previous flare
$x_i=E_{i-1}$ or the the energy ratio between successive flares, 
$x_i=E_{i}/E_{i-1}$. The quantity
$P(x_i>X \vert \Delta t_i <T)$ is given by 
$
P(x_i>X \vert \Delta
 t_i<T) \equiv \frac{N(X,T)}{N(T)}$,  where $N(X,T)$ is the number of couples of subsequent events
fulfilling both  $x_i>X$ and $\Delta t_i <T$, whereas $N(T)$ is the
 number of couples satisfying only the condition $\Delta t_i <T$.
\begin{figure}
    \centering
\vskip+0.5cm
\includegraphics[width=7.cm]{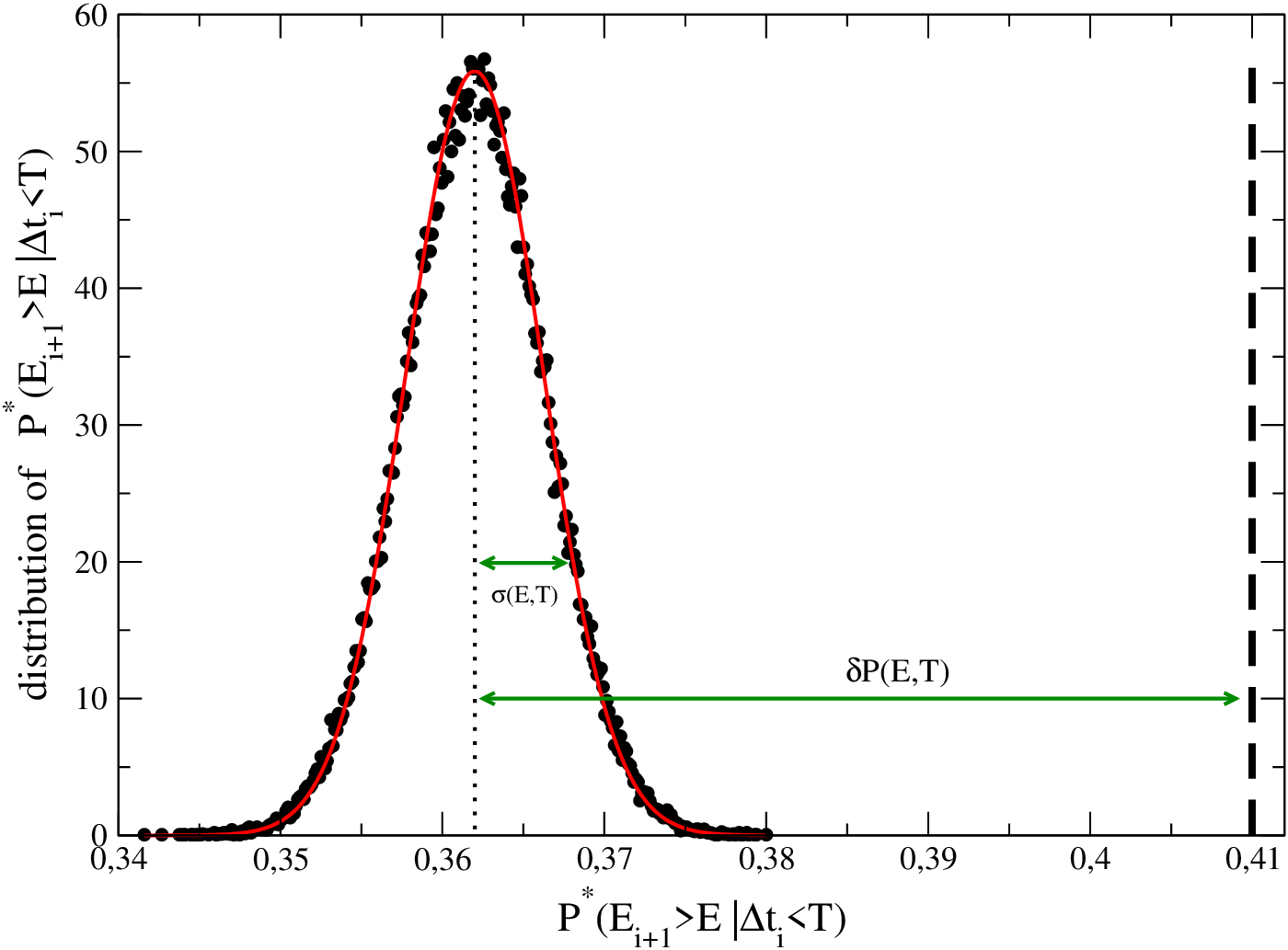}
%\includegraphics[scale=.42]{fig1.eps}
%    \rotatebox{0}{\resizebox{.65\textwidth}{!}{\includegraphics{fig1.eps}}}
  
%\includegraphics[width=8cm]{fig1b.eps}
%    \rotatebox{0}{\resizebox{.65\textwidth}{!}{\includegraphics{fig1.eps}}}
%\includegraphics[width=8cm]{fig1c.eps}
%    \rotatebox{0}{\resizebox{.65\textwidth}{!}{\includegraphics{fig1.eps}}}
    \caption{ Distribution of $P^*(E_i>E \vert
    \Delta t_i<T)$ for $E=3 E_0$, $T=1$h and $10^5$  realizations 
of the reshuffled catalog represented by black circles. The continuous
    red line is the fit with a Gaussian distribution with an average value of 
0.3619 (dotted line) and a standard deviation
    $\sigma (E,T)=6.5 \ \ 10^{-3}$. The dashed line indicates the value 
$P(E_i>E \vert
    \Delta t_i<T)=0.4098$ obtained in the real catalog. 
%(b) The
%    distribution $\delta P(m_i>M \vert \Delta t_i<T)$ as function of
%    $M$ for different values of $T$. For each data point $(M,T)$ the error
%    bar is the standard deviation $\sigma(M,T)$. Orange left triangles are
%    for $T=1$h restricting only to flares with $\Delta t_i >2
%    (t^f_{i-1}-t^0_{i-1})$. (c)  The
%    distribution $\delta P(m_{i} = M \vert \Delta t_i<T)$ as function of
%    $M$ for different values of $T$. 
} 
    \label{fig1}
\end{figure}
 In order to detect eventual correlations in the catalog, we introduced 
the probability difference $\delta P(x_i>X \vert \Delta t_i<T) 
\equiv  P(x_i>X \vert \Delta t_i<T)-P(x_i>X)$, where $P(x_i>X)$ is the
unconditional probability to have $x_i>X$. A positive (negative) $\delta P$
indicates a higher (lower) probability to find flares with
$x_i>X$ if one restricts the analysis to couples with $\Delta t_i>T$. Hence,  values of
$\delta P$ different from zero  suggest
the existence of correlations between $x_i$ and $\Delta
t_i$. Evidently, because of statistical fluctuations,  
$\delta P$ is never exactly equal to zero even in catalogs where $x_i$
and $\Delta t_i$ are uncorrelated. In order to explicitly take into
account the role of statistical fluctuations, we 
computed the quantity $P^*(x_i>X \vert \Delta t_i<T)$, 
 defined as $P$, but in a catalog where flare energies were
 randomly reshuffled. The comparison with thereshuffled catalog is the basis
of the surrogate data technique (Theiller et al. \cite{The}, Schreiber \& Schmitz \cite{Sch}), 
recently applied to investigate coherent structures in  space plasma
(Sahraoui \cite{Sah}).  In the reshuffled catalog  $x_i$ was by
 construction  uncorrelated to $\Delta t_i$ and 
$P^*(x_i>X \vert \Delta t_i<T)$ fluctuated around its
average value $P(x_i>X)$. The amplitude of these fluctuations defined
the significance level, $\sigma (X,T)$, which allowed us to distiguish between 
the presence and absence of  correlations. 
The method is schematically presented in Fig.1 for $x_i=E_i$, $X=E=3 E_0$
 and $T=1$ h. 
$P^*(E_i>E \vert \Delta t_i<T)$ takes different values  for each
realization of the reshuffled catalog. 
We produced $10^5$ independent realizations of the catalog with reshuffled
energies and observed that
$P^*(E_i>E \vert \Delta t_i<T)$  is
 Gaussian-distributed with a mean $P(E_i>E)$ and a standard deviation
$\sigma (E,T)$.  
Similar results were obtained for other values of $E$
 and $T$ and for the other  definitions of $x_i$. 
Therefore,  $\vert\delta P(x_i>X \vert \Delta t_i<T)\vert>\sigma (X,T)$  indicates
that energies in the real catalog follow a significantly different 
organization 
than events in the reshuffled catalog. 
%In the specific case of Fig.1, we find that 
%$P(E_i>E \vert \Delta t_i <T)$ for the real catalog is
%larger than all values obtained in reshuffled catalogs. This implies that 
% the number of flares with energy larger than $E$, after small
% waiting times, is greater in the real than in a catalog with no correlations.  

\begin{figure}
    \centering
\vskip+0.5cm
\includegraphics[width=7.cm]{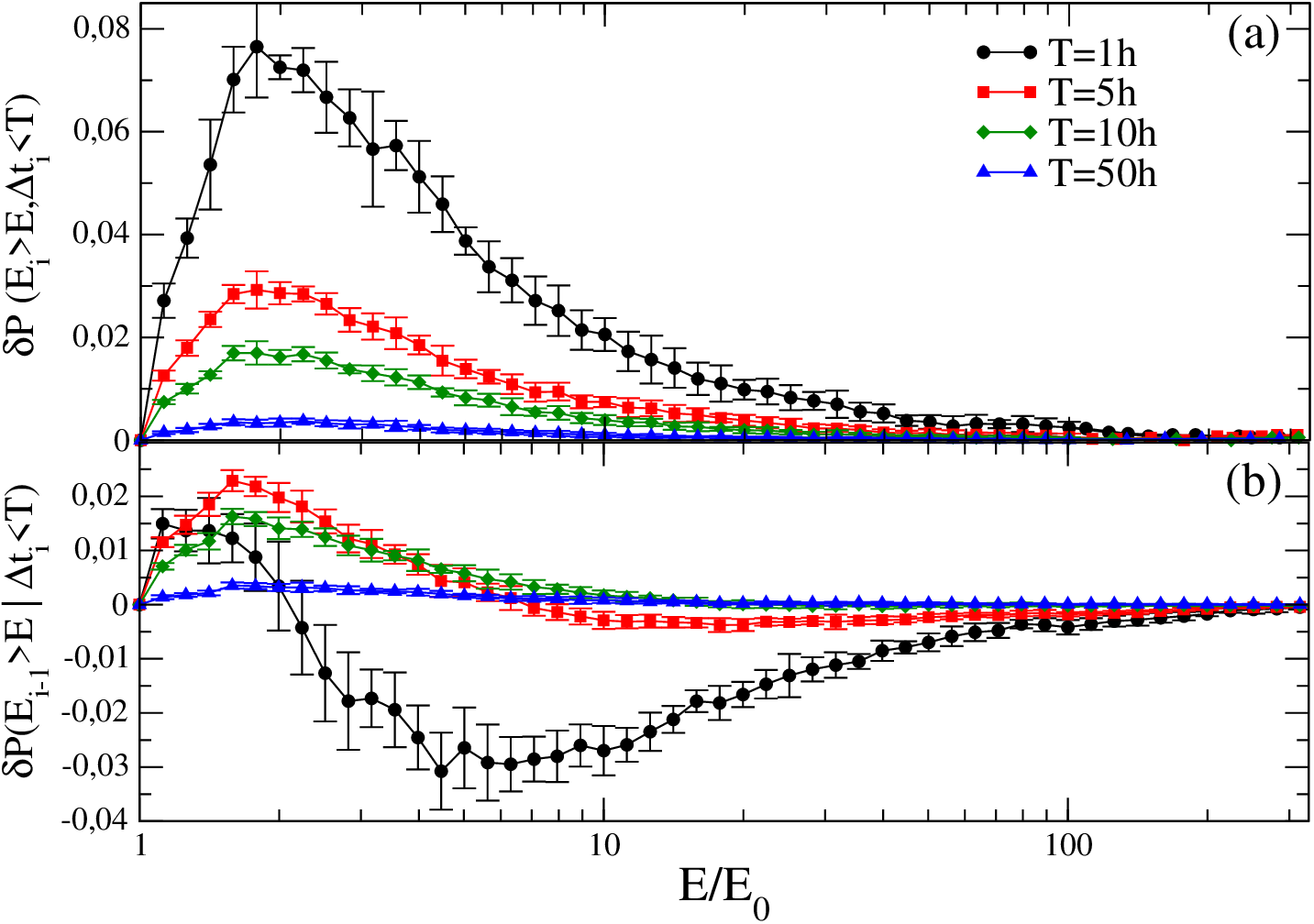}
\caption{ Probability difference
 $\delta P(E_i>E \vert \Delta t_i<T)$ (a) and
$\delta P(E_{i-1}>E \vert \Delta t_i<T)$ (b) as a function of
    $E/E_0$ for different values of $T$. For each data point the error
    bar is the standard deviation $\sigma(E,T)$.}
\label{fig2}
\end{figure}
\begin{figure}
  \vskip0.35cm
    \centering
\includegraphics[width=7.cm]{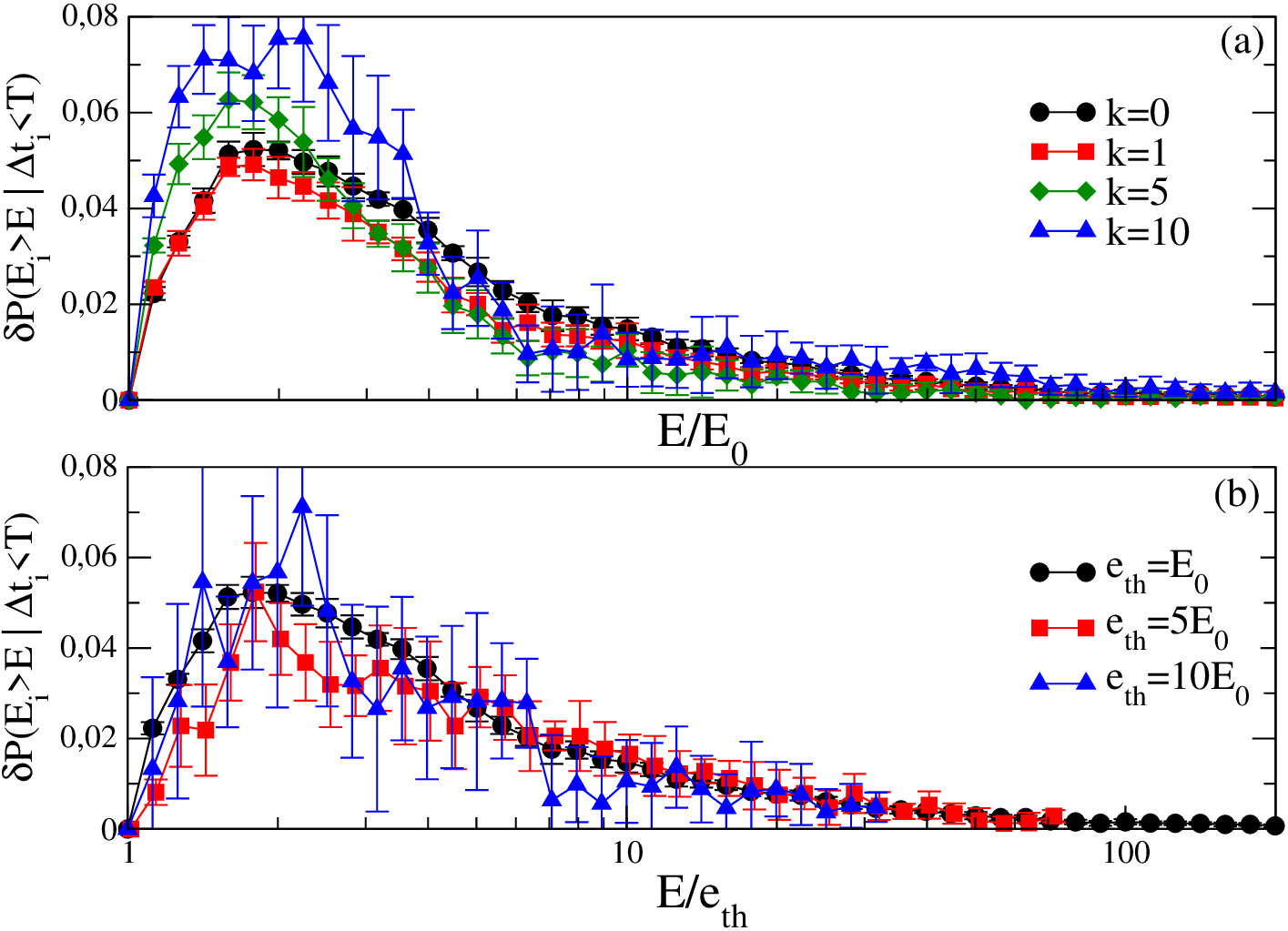}
\caption{(a) Probability difference
$\delta P(E_{i-1}>E \vert \Delta t_i<T)$ as a function of
$E/E_0$ for $T=1$h and different $k$ values. (b) Same quantity
as function of $E/e_{th}$ for $T=1$h and different values of $e_{th}$.
For each data point the error bar is the standard deviation
$\sigma(E,T)$.}
\label{fig3}
\end{figure}

\section{Results}

In Fig.2a we plot $\delta P(E_i>E \vert \Delta t_i<T)$ and in Fig.2b 
$\delta P(E_{i-1}>E \vert \Delta t_i<T)$ for different values    
of $T$.  The standard deviation for each data point $\sigma (E,T)$ is
represented as the error bar. 
In Fig.2a we notice that $\delta P(E_i>E \vert \Delta t_i<T)$ takes 
always positive values beyond error bars. This implies that in the
real catalog  the number of couples 
fulfilling both conditions is  greater than
in catalogs where energies and intertimes are
uncorrelated. More precisely we find that  
for each  given value of $E$, 
 $\delta P(E_i>E \vert \Delta t_i<T)$ decreases by
increasing $T$. This implies that 
the probability to find flare couples with
 the second flare energy higher than $E$ 
decreases if one includes events with larger 
$\Delta t$ in the analysis. 
This result is disagrees with what is
expected according to the RV model, which predicts larger flares
after longer waiting times. On the other hand, obscuration can
represent a possible explanation of the above result. Indeed,
according to the selection procedure, for flares close in time, the
second event is recorded in the catalog only if it produces an
increase in the flux of at least  $40\%$ of the level of the
previous flare. Moreover, the light curve of large flares may hide 
smaller flares close in time.
Obscuration effects then reduce the
probability to find a small flare after a short waiting time and may
introduce spurious correlations between waiting time and the successive
flare energy. 

We then explicitly explored the role of obscuration on
$\delta P(E_i>E \vert \Delta t_i<T)$.   
As a  first analysis we considered 
only flare couples with a temporal distance
greater than $k$ times  the duration of the first flare. 
Spurious effects due to obscuration should disappear by increasing
$k$. We observed (Fig.3a) that this extra condition did not reduce
correlations  between waiting
times and energies, since they are still present also for the highest values of
$k$. To further prove that correlations cannot be attributed to
spurious effects related to obscuration, we performed 
the same analysis as in Fig.1a, setting different lower energy thresholds
$e_{th}$. Indeed, by increasing 
the value of the minimum energy required for flares to be included 
in the analysis, the percentage of flares hidden by obscuration 
was reduced and obscuration effects should progressively decrease.
In Fig. 3b  we plot the results for different energy thresholds $e_{th}$ as a
function of $E/e_{th}$. No significant dependence on
$e_{th}$ is detected, indicating that correlations cannot be attributed to
obscuration effects. We conclude that for 
couples of events close in time it is more probable to have
the second event with a high energy than 
for events distant in time.
\begin{figure}
    \centering
\vskip+0.75cm
\includegraphics[width=7.cm]{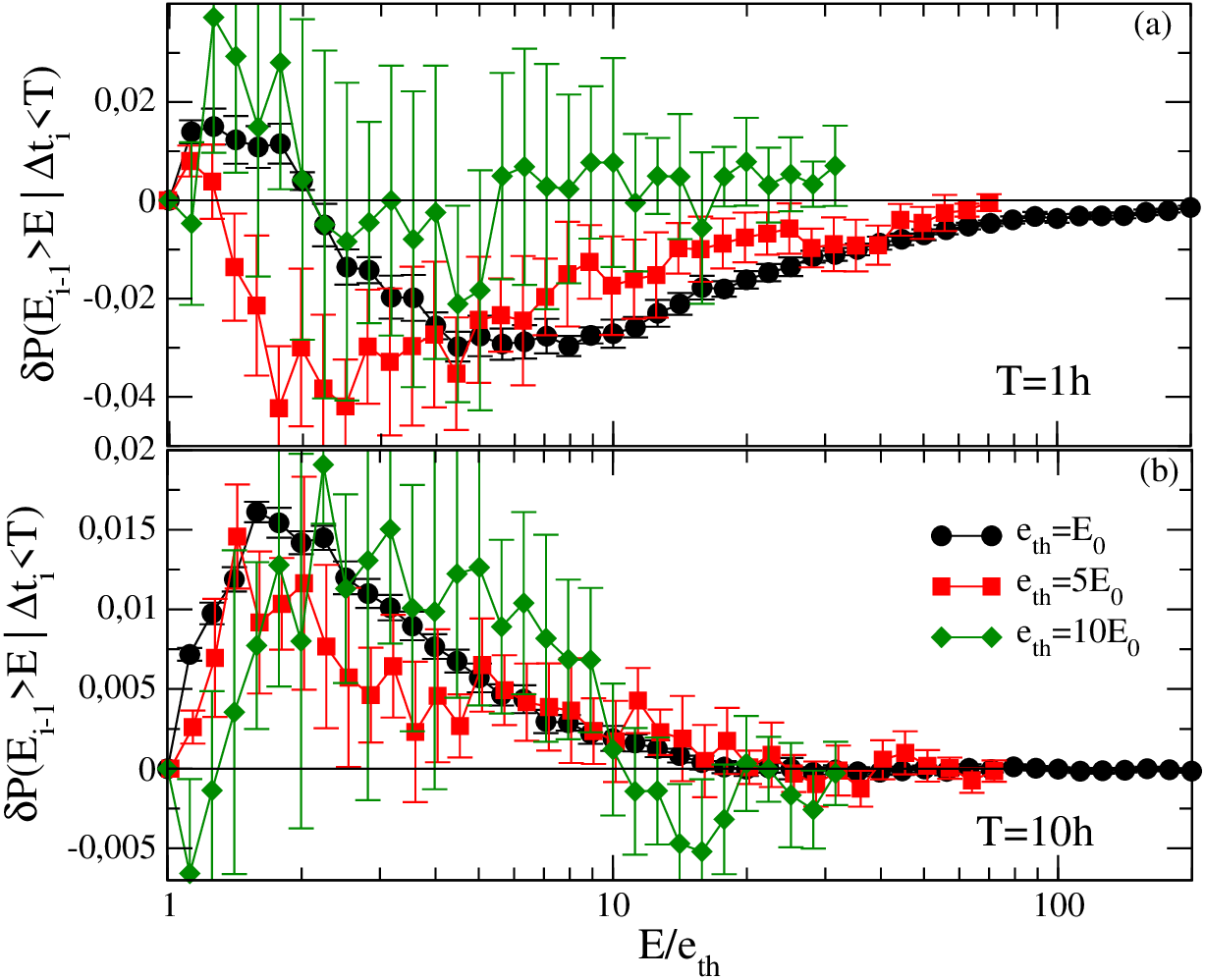}
\caption{ Probability difference
    $\delta P(E_{i-1}>E \vert \Delta t_i<T)$ as a function of
    $E/e_{th}$ for $T=1$h (a) and $T=10$h (b) for different values of $e_{th}$.
For each data point the error bar is the standard deviation $\sigma(E,T)$.}
\vskip0.5cm
\label{fig4}
\end{figure}

\begin{figure}
    \centering
\includegraphics[width=7.cm]{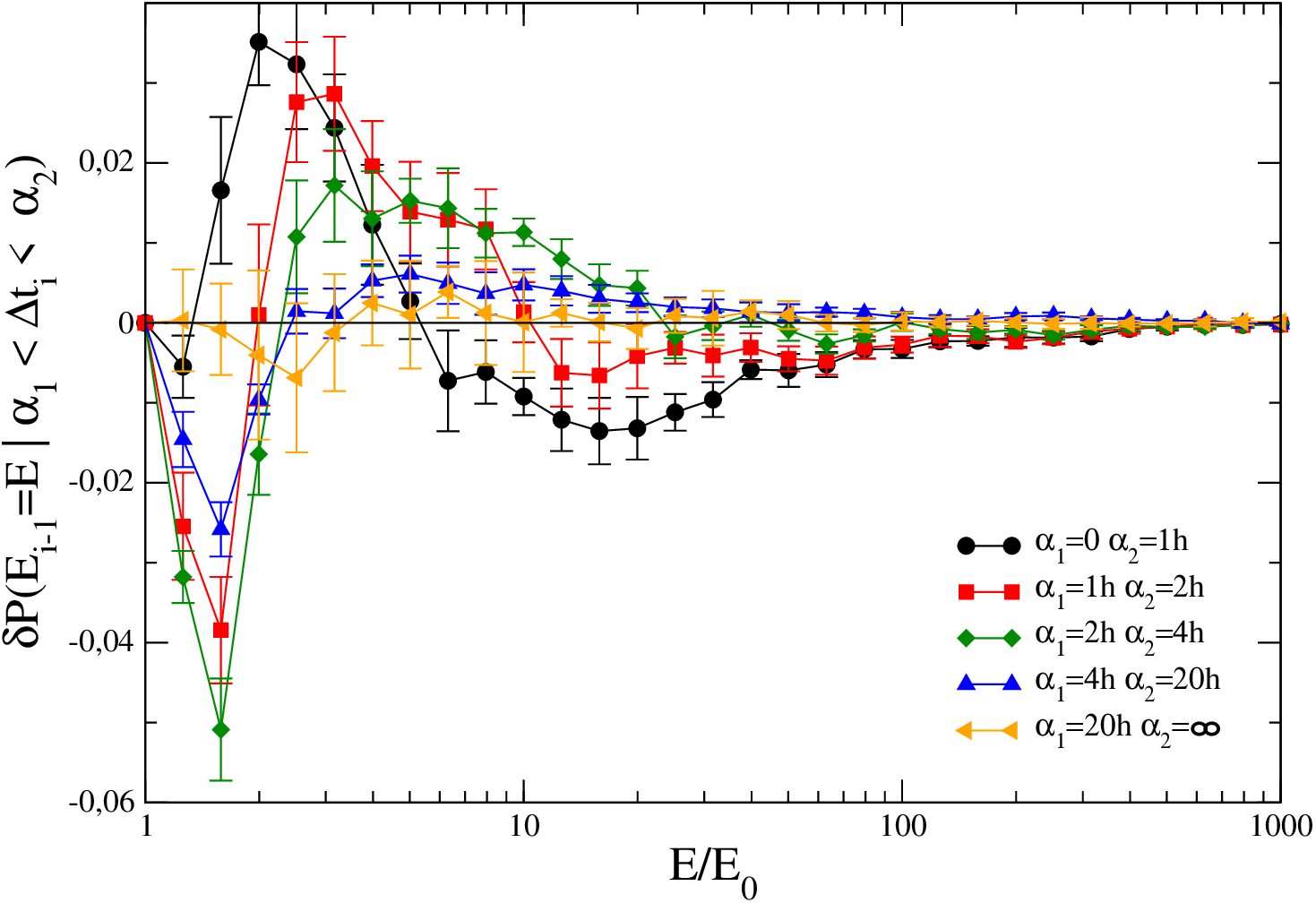}
\caption{Probability density difference
$\delta p(E_{i-1}=E \vert \alpha_1<\Delta t_i<\alpha_2)$ 
versus $E/E_0$ for different values of $\alpha_1$ and $\alpha_2$. 
}
\label{fig5}
\end{figure}
In Fig.2b we performed the same analysis as in Fig.2a for the
correlations between the waiting time and the previous flare
energy. In contrast to Fig.2a the quantity 
$\delta P(E_{i-1}>E \vert \Delta t_i<T)$ is not always positive now. In
particular for $T=1$h, $\delta P(E_{i-1}>E \vert
\Delta t_i<T)$ is positive at small $E$ and negative at large
$E$. In order to  verify if the behavior at small $T$ is affected
by obscuration, we performed the same analysis as in Fig.3b, 
imposing the condition on the lower energy threshold $e_{th}$. 
In Fig.4a we present $ \delta P(E_{i-1}>E \vert \Delta t_i<T)$
for $T=1$h and different energy thresholds. We found that data strongly
depend on $e_{th}$, in particular for $e_{th}=10E_0$, $ \delta
P(E_{i-1}>E \vert \Delta t_i<T)$ is 
%always positive and 
comparable
with the statistical fluctuations. This analysis suggests that the correlations
observed at small $\Delta t$ are related to obscuration
effects. Conversely, data for $T=10$h are not significantly
affected by $e_{th}$ and $ \delta P(E_{i-1}>E \vert \Delta t_i<T)$
always takes positive values beyond error bars (Fig.4b). This result indicates that 
correlations detected for events with temporal distances between $1$h
and $10$h are not significantly affected by obscuration.
% that infact predicts an opposite trend.        
We conclude that couples of events distant in time between $1$h and $10$h
 tend to have the first event with a higher energy than the one expected
for a process where time and energy are uncorellated.

To obtain more detailed information on correlations between $E_{i-1}$ and
$\Delta t$, we considered 
the quantity $dp(E,\alpha_1,\alpha_2)\equiv \delta P(E_{i-1}=E \vert
\alpha_1<\Delta t_i<\alpha_2)$, 
i.e. the difference between  conditional and unconditional
  probability densities to have $E_{i-1}=E$ 
if $\Delta t_i \in ]\alpha_1,\alpha_2[$.
This quantity can be obtained
 deriving the cumulated probability $\delta P(E_{i-1}>E \vert
 \alpha_1<\Delta t_i<\alpha_2)$ with respect to $E$. 
Being a derivative, it is affected by higher fluctuations. 
For every choice of $\alpha_1$, the value of $\alpha_2$ was chosen 
to have the same number of events, about $9000$, for all 
 curves. We noticed that for $\Delta
 t>\alpha_1=20$h the difference in the probability density $dp$ was
 very close to zero, indicating the absence of correlations.     

We now discuss the results of Fig.5 at fixed $E/E_0$ and for varying $\alpha_1$.
 We first observed that for small of $E$, 
 $dp$ is always negative for all values of $\alpha_1$. 
The negative
 region is restricted to very small values of $E\lesssim 1.5E_0$ for
 $\alpha_1=0$ and increases to  $E \lesssim 2.3E_0$ for  larger
 $\alpha_1$. Within this interval $dp$ is non monotonic in $\alpha_1$,
 in particular, 
it reaches the minimum value for $\alpha_1=2$h and then tends to
 zero.  For $\Delta t_i$ in a given range, negative values of $dp$ imply that 
it is less probable to have a flare with an energy $E_i=E$ compared
to the case where $E_i$ and $\Delta t_i$ are uncorrelated.
 Results in Fig.5 then
imply that it is improbable to have small $\Delta t$ after
 very small flares and that these anticorrelations vanish for  large
 $\Delta t$.  
In particular the highest negative values of $dp$ are obtained 
for $\Delta
 t \in [2h,4h]$  after an $E \lesssim 2E_0$ flare. 

If these anticorrelations were
a spurious effect of obscuration, we should rather observe a different
behavior.   
Indeed, energy reshuffling produces more events shortly after a large
flare compared to the real catalog. Consequently, 
we should observe an excess, and not a deficit, of events
after a small flare in the real catalog compared to the reshuffled one 
because the number of events is conserved.
 We now discuss higher values of
 $E \gtrsim 10E_0$. In this case  $dp$ moves from negative values for
 $\alpha_1=0$ and $\alpha_1=1$h to positive values for $\alpha_1=2$h
 and $\alpha_1=4$h and then tends to zero for higher values of
 $\alpha_1$. This indicates that there is 
 a deficit of events with $\Delta t<2$h after energetic flares, whereas for
  $\Delta t \in $[2h,4h] the number of observed flares is
 larger than the one expected for an uncorrelated process.
 The above results suggest that after a large flare a
 recovery time of about $2$ hours takes place, when only a small number of flares
 is observed. After this time the number of
 observed flares reaches a maximum value for $\Delta t \lesssim 4$h and
 then decreases to the background level for larger $\Delta t$. 
According to the dependence on $e_{th}$ observed in Fig.4a, the
recovery time can be a spurious effect related to obscuration.

The number of events $n(t)$ occurring at the time $t$ after an
energetic 
flare can be
explicitly evaluated  following the method of Lippiello (\cite{Lip2}).
More precisely, we define as
a ``main'' flare  an event with the energy $E\ge E_{main}=10E_0$. Indicating with
$t_i$ the occurrence times of main-flares we compute the
quantity
${\cal N}(t,E_{main}) = \sum_{i} n(t-t_i)\Theta(t_{i+1}-t) 
$
, where $\Theta(x)$ is the Heaviside step function, and the sum extends
over all main-flares. 
Here $n(t-t_i)$ is the number of flares with an energy lower than $10E_0$ 
occurring  during the time $t-t_i$ after the
i-th main flare, and the sum extends over all main-flares  in the
catalog. 
Assuming that $n(t-t_i)$ is
time translationally invariant we obtain
\be
n(t)=\frac{{\cal N}(t,E_{main})}{\sum_{i} \Theta(t_{i+1}-t)},
\label{6}
\ee
which represents the average number of flares occurring in a period of
a duration $t$  after a mainflare. The data (Fig.6a) agree with the
previous results,  with a maximum number of flares at  $t\simeq 4$h, 
decaying at longer times. Notice that the asymptotic decay is
consistent with a power law $t^{-p}$ with $p\simeq 1$, reminiscent of
the Omori law for seismic sequences (Omori \cite{Omo}, de Arcangelis et
al \cite{deA}). 
\begin{figure}
    \centering
\vskip+0.6cm
\includegraphics[width=7.cm]{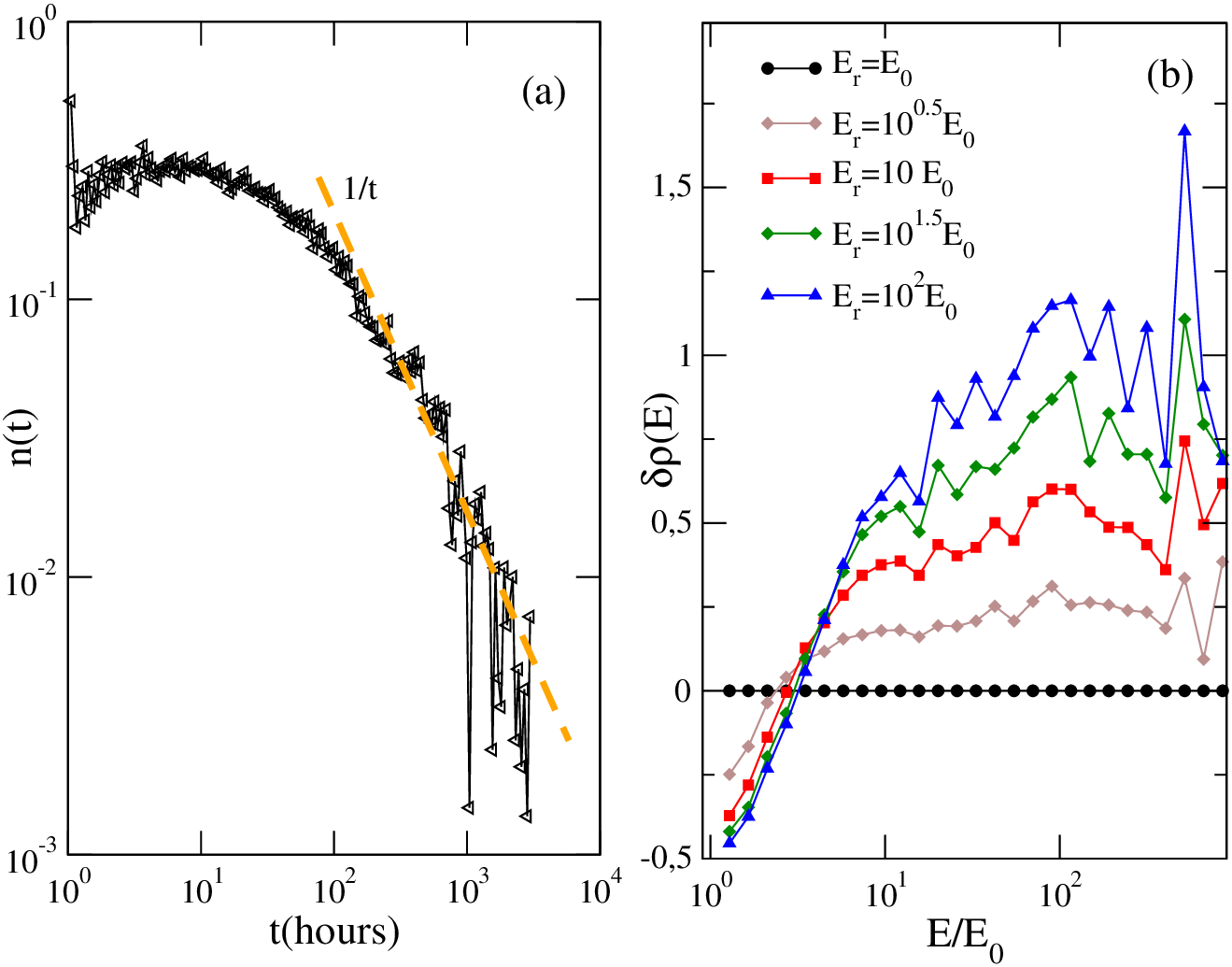}
%    \rotatebox{0}{\resizebox{.65\textwidth}{!}{\includegraphics{fig1.eps}}}
    \caption{ (a) Average number of flares $n(t)$ as a function
of the time $t$ occurring after a main
flare with an energy $E \ge E_{main}=10E_0$. The data are averaged over $4057$ main flares.
(b) The difference $\delta\rho (E)$ between the energy distribution for the
first $m=20$ flares occurring after an event with a higher energy than $E_r$ and
the entire catalog energy distribution and different $E_r$ values.
    }
    \label{fig6}
\end{figure}

We now  discuss the existence of
correlations between energies of subsequent flares. In order to do so,
we computed the energy distribution  $\rho(E)$ of the first $m$ flares occurring
after a flare with a larger energy than a reference value
$E_r$. 
Since the analysis is restricted  to all flares with $E \ge E_0$, obviously
$\rho(E)$ coincides with   
the flare energy distribution of the whole catalog $\rho_T(E)$, for $E_r=E_0$. 
In Fig.6b we plot $\delta \rho (E)=\rho(E)-\rho_T(E)$ for
$m=20$ and different values of $E_r$. Similar results are
obtained for other values of $m$. 
Deviations from $\rho_T(E)$ become more and more evident for
increasing values of $E_r$. In particular, the higher $E_r$ is,
the larger is the probability to have subsequent flares with higher
energy and the lower is the probability to have small flares. 

In order to get further insights in
 the correlations between flare energies, we computed the quantity
$\delta p(E_{i}=\lambda E_{i-1}
\vert \Delta t_i<T)$, i.e. the difference between the conditional and the
unconditional 
probability density to have the energy of the subsequent flare $\lambda$ times
the previous one.
We found (Fig.7a) that for $T=1$h, $\delta
p(E_{i}=\lambda E_{i-1}
\vert \Delta t_i<T)$ is significantly different from zero for
all $\lambda$ values. This indicates that the energies of two
subsequent flares are correlated. Moreover we observed that 
these correlations depend on
the time separation between the two flares and are practically zero for
$\Delta t_i>10$h (Fig.7a). 
We verified that this result is not affected by obscuration  performing
the same analysis for $T=1$h and different lower energy thresholds $e_{th}$
(Fig.7b). We found that curves for different $e_{th}$ coincide within
statistical fluctuations. This indicates that energy correlations and
their dependence on time separation are a physical property, not a
spurious effect due to obscuration.
Curves present a maximum for $\lambda \gtrsim 1$, indicating that it
 is more probable to find the next flare with an energy close to
 but  slightly higher than the previous one.

\begin{figure}
    \centering
\vskip+0.5cm
\includegraphics[width=7.cm]{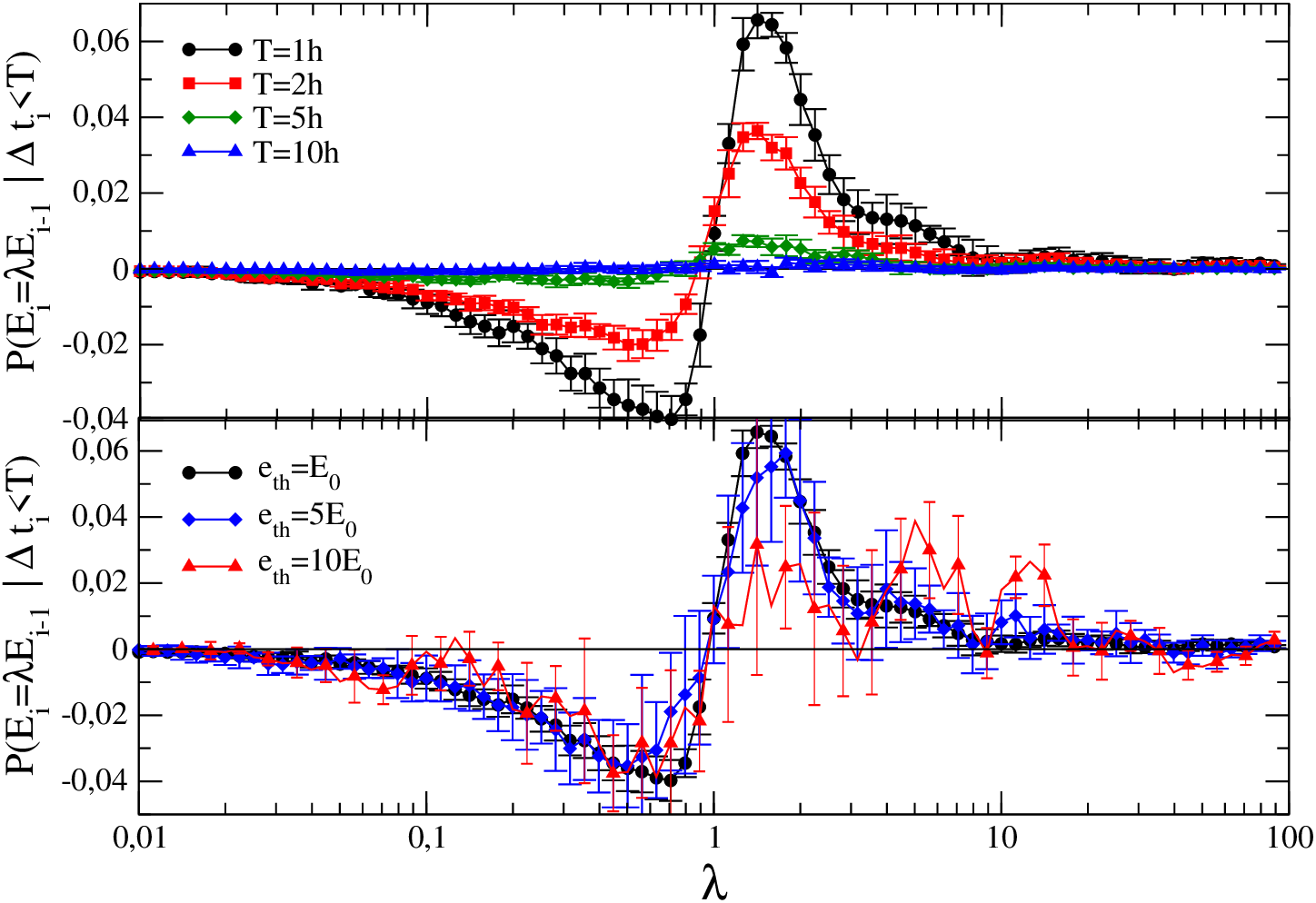}
%    \rotatebox{0}{\resizebox{.65\textwidth}{!}{\includegraphics{fig1.eps}}}
    \caption{Probability density difference
$\delta p(E_i=\lambda E_{i-1} \vert \Delta t_i<T)$ as a function of
    $\lambda$ for different $T$ (a) and for $T=1$h and different
      $e_{th}$ (b).}
    \label{fig7}
\end{figure}
\section{Conclusions}
In conclusion, we presented a statistical analysis 
of the  GOES catalog 
indicating the existence of
time-energy correlations between successive events not to be attributed to
obscuration effects. 
More precisely, we observed that for couples of events close in time ($T<1$h),
the second event tends to have a high energy. Moreover couples of events
distant in time between 1h and 10h have the first event with a high energy.
%A detailed analysis of the dependence of these correlations on the 
%range of the temporal distance, confirms that after energetic flares the number
%of observed events occurring between 2h and 4h later is larger than in an
%uncorrelated catalog. Indeed 
The analysis of the rate decay after 
large flares shows evidence that the largest number of events is detected about 4h after 
the occurrence of the main event. Finally the distribution of flare energies
confirms that the higher the flare energy, the larger the number of
subsequent events with high energy.
The existence of time-energy correlations suggests the possibility of 
scaling laws relating time with the energy released in a flare. This is a still open
question, which could provide interesting insights in the energy storage and
release mechanisms at the origin of solar flare occurrence.

%{\small Acknowledgments. 
%This research was supported by EU Network Number
%MRTN-CT-2003-504712, MIUR-PRIN 2004, MIUR-FIRB 2001.}
%089958642

\end{document}